# Superconductivity in the non-oxide Perovskite MgCNi$_3$

T. He[1], Q. Huang[2], A.P. Ramirez[3], Y. Wang[4], K.A. Regan[1], N. Rogado[1], M.A. Hayward[1], M. K. Haas[1], J.S. Slusky[1], K. Inumaru[1], H.W. Zandbergen[1], N.P. Ong[4], and R.J. Cava[1]

[1]*Department of Chemistry and Princeton Materials Institute, Princeton University, Princeton NJ,*
[2]*Department of Materials and Nuclear Engineering, University of Maryland, College Park, MD, and NIST Center for Neutron Research, Gaithersburg, MD,*
[3]*Condensed Matter and Thermal Physics Group, Los Alamos National Laboratory, Los Alamos NM,*
[4]*Department of Physics, Princeton University, Princeton NJ*

(March 14, 2001)

The oxide perovskites are a large family of materials with many important physical properties. Of particular interest has been the fact that this structure type provides an excellent structural framework for the existence of superconductivity. The high $T_c$ copper oxides are the most famous examples of superconducting perovskites, but there are many others [1]. Intermetallic compounds have been the source of many superconducting materials in the past, but they have been eclipsed in recent years by the perovskite oxides. The recent discovery of superconductivity in MgB$_2$ [2] suggests that intermetallic compounds with simple structure types are worth serious reconsideration as sources of new superconducting materials. Here we report the observation of superconductivity at 8 K in the perovskite structure intermetallic compound MgCNi$_3$, linking what appear at first sight to be mutually exclusive classes of superconducting materials. The observation of superconductivity in MgCNi$_3$ indicates that MgB$_2$ will not be the only one of its kind within the chemical paradigm that it suggests for new superconducting materials.

The variable stoichiometry compound MgC$_x$Ni$_3$, for 0.5>x>1.25 has been previously reported, and assigned to the perovskite structure type by analogy [3,4]. Neither its crystal structure nor its physical properties had been determined previously. In this study, samples with nominal formula MgC$_x$Ni$_3$ for x=1.5, 1.25, 1.1, 1.0 and 0.9 were prepared. Starting materials were bright Mg flakes (Aldrich Chemical), fine Ni powder (99.9% Johnson Matthey), and glassy carbon spherical powder (Alfa AESAR). Starting materials were mixed in half-gram batches, and pressed into pellets. The pellets were placed on Ta foil, which was, in turn, placed on an Al$_2$O$_3$ boat, and fired in a quartz tube furnace under a mixed gas of 95% Ar 5% H$_2$. The samples were heated for half an hour at 600°C, followed by one hour at 900°C. After cooling, they were ground, pressed into pellets, and heated for an additional hour at 900°C. Due to the volatility of Mg encountered during the synthesis of this compound, 20% Mg in excess of the stoichiometric ratio was employed in the initial mixtures. The structural determination, described below, indicated that the compound formed was stoichiometric in metals and that no excess Mg remained in the samples. The larger batch of material (3 grams) employed in the neutron diffraction study was synthesized by the same method.

The crystal structure of a superconducting sample ($T_c \cong$ 7.3K, determined magnetically, as described below) of nominal composition MgC$_{1.25}$Ni$_3$ was determined by powder neutron diffraction at ambient temperature [5]. The observed powder diffraction pattern and the pattern calculated for the final structural model are shown in figure 1. The formula for the superconducting phase was found to be MgC$_{0.96}$Ni$_3$. The compound has the classical cubic perovskite structure, space group Pm-3m, with a=3.81221(5) A. The positions for the atoms are: Mg: 1a (0,0,0), C: 1b (0.5,0.5,0.5) and Ni: 3c (0, 0.5,0.5). The temperature factors are 0.90(3), 0.54(4), and 0.75(1) A$^2$ for Mg, C, and Ni, respectively. Refinements were performed with variable stoichiometry allowed for the C site. The C site occupancy was found to be 0.960(8) making the exact stoichiometry MgC$_{0.96}$Ni$_3$. In agreement with what is expected from the nominal composition, a small amount of unreacted graphite (2 weight %) was found in the sample. The sensitivity of the superconductivity to the C content of the perovskite phase made it necessary for carbon excess to be added to the initial mixtures to insure attainment of the superconducting composition. The refinement agreement, weighted profile agreement, and chi-squared values obtained were R=5.14%, R$_{wp}$=6.39% and $\chi^2$= 1.258, indicating the high quality of the structural model. The perovskite crystal structure for MgCNi$_3$ is shown in the inset to figure 1. Comparison to a familiar oxide perovskite such as CaTiO$_3$, for example, indicates the structural equivalencies between Ca and Mg, Ti and C, and O and Ni.

The temperature dependent magnetization at low temperatures was measured for these materials, in the form of loose powders, in a Quantum Design PPMS magnetometer under an applied DC field of 15 Oe. The data, taken on heating after cooling in the absence of a field, are presented in figure 2. The data show that the magnetic onset for the superconducting transition ranges between 7.1 K for a nominal carbon content of 1.1 per MgC$_x$Ni$_3$ to 7.4 K for nominal carbon content x = 1.5. The superconducting transition turns off abruptly for nominal C contents between x = 1.1 and x = 1.0. All samples in the range of carbon contents shown appear single phase by powder X-ray diffraction (which would not be sensitive to the presence of graphite impurity). Their crystallographic cell parameters vary smoothly between 3.805(1) A at nominal x=1.0 and 3.816(1) A at x = 1.5 due to a difference in carbon content of the perovskite phase. The neutron diffraction results indicate that the composition of the superconducting phase for nominal x=1.25 is MgC$_{0.96}$Ni$_3$. Therefore the data suggest that the highest T$_c$ in the perovskite is obtained for the perfectly stoichiometric composition MgC$_{1.0}$Ni$_3$, which requires the presence of excess carbon to form under our synthetic conditions. Further, the data indicate that the superconductivity disappears abruptly for carbon contents between 0.96 and approximately 0.90 per formula unit. The reason for this is not yet known.

The temperature dependent resistivity for a polycrystalline pellet of nominal composition MgC$_{1.5}$Ni$_3$ made from powder first reacted as described and then pressed at a pressure of 10 Kbar for one hour at 700 C is shown in figure 3. The normal state resistivity at room temperature is relatively low, even for the polycrystalline sample, about 90 μohm-cm. The resistivity does not decrease much with temperature, with a resistivity ratio of approximately 2.1 to 9K. The temperature dependence is consistent with what is expected for a poor metal, but whether the shape of ρ(T) is intrinsic or not remains to be verified by future work on single crystals if they become available. The inset shows a detail of the resistive superconducting transition. The transition is very sharp, indicating that the fraction of material with the optimum carbon stoichiometry for the highest T$_c$ is above the percolation threshold. The midpoint of the resistive transition is 8.4 K, the 90-10% transition width is 0.1K, and the resistive onset temperature is 8.5 K.

Specific heat measurements were made using a relaxation method in a commercial calorimeter (Quantum Design). The powder sample was cold-sintered with Ag powder in a MgCNi$_3$: Ag ratio of 3:2 by mass. Fig. 4a shows the specific heat divided by temperature, C(T)/T versus T, with the contribution from Ag subtracted. This figure allows us to perform an equal-area extrapolation to determine both the midpoint of the thermodynamic transition, as well as the jump, ΔC/T$_c$, assuming a mean-field transition. The transition midpoint is 6.2K, significantly below the resistive onset. This value, as well as the breadth of the jump, ~1.7K, is likely due to a small carbon stoichiometry variation in the sample: the transition width implies that the sample studied possessed a variation of ~ 0.01 in C content. We find ΔC/T$_c$ ~ 19mJ/moleNi-K$^2$.

In Fig.4b we plot C(T)/T versus T$^2$. Assuming that C(T) has two leading contributions, a linear Fermi-liquid term, γT, where γ is the Sommerfeld constant, and a cubic lattice term, this plot allows an extrapolation of the normal-state behavior to T=0 where γ is determined as the T = 0 value of C(T)/T. By this method, we find that γ ~ 10 mJ/moleNi-K$^2$. If we assume that superconductivity is mediated by phonons, then ΔC/T$_c$ is related to γ by a numerical solution of the Eliashberg equations [6,7]:

$$\Delta C/T_c = (1.43 + 0.942\lambda_{ph}^2 - 0.195\lambda_{ph}^3)\gamma,$$

where $\lambda_{ph}$ is the electron-phonon coupling constant. From the above data we ascertain ΔC/T$_c$/γ = 1.9 ± 0.1, which leads to $\lambda_{ph} = 0.77^{+0.17}_{-0.09}$. This value of $\lambda_{ph}$ is in the range of conventional phonon-mediated pairing, though given the large amount of Ni in the compound, it is too early to attribute the superconductivity to a conventional microscopic mechanism.

Although the structural analogy between superconducting intermetallic perovskite MgCNi$_3$ and superconducting oxide perovskites like (Ba,K)BiO$_3$ [1] is transparent, the electronic analogy is at first sight not so obvious. A strict electronic analogy is also possible, however, between these seemingly completely different materials. For the oxide perovskites, an important characteristic of the superconductivity is that the electronic states at the Fermi energy involve holes in the oxygen electronic orbitals. The Ni in the MgCNi$_3$ perovskite, which is structurally equivalent to O in the oxides, may also be found to have holes in its electronic orbitals at the Fermi energy. Due to the nearly filled *d* orbitals in elemental nickel, intermetallic compounds of Ni, especially in the presence of electropositive elements such as Mg, often have filled or nearly completely filled Ni *d* states. If this is the case for MgCNi$_3$ then the conduction will involve holes in Ni *d*-states, in a strict electronic analogy to the holes in the O *p*-states in the perovskite oxide superconductors. The structural and electronic equivalence of the oxides and intermetallic perovskite superconductor MgCNi$_3$ would then be complete. The fact that superconductivity rather than ferromagnetism occurs in a compound where so much nickel is present is one of the surprises that this material has to offer for further consideration.

### Acknowledgments
This research was supported by grants from the U.S. National Science Foundation and the U.S. Department of Energy.

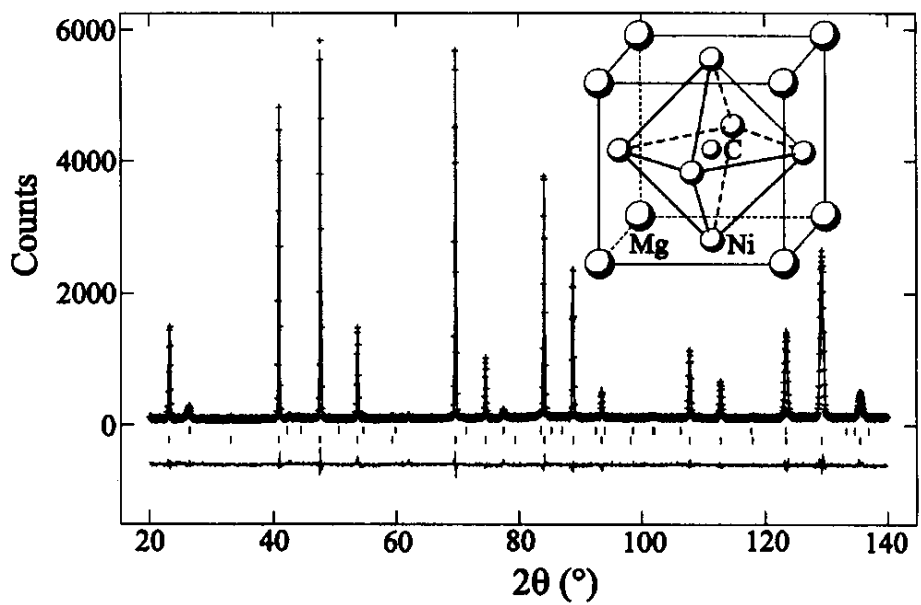

Fig. 1. The powder neutron diffraction pattern for the sample of nominal composition MgC$_{1.25}$Ni$_3$. The perovskite crystal structure for the superconducting compound MgCNi$_3$ is shown as an inset. Data are shown as crosses, and the difference plot between model and data shown directly below. The vertical lines (bottom) show the Bragg peak positions for the MgCNi$_3$ phase. The sample contains 2 weight % graphite (about 25 mole %) in agreement with the nominal composition. Positions of the graphite peaks are shown as vertical lines above those for MgCNi$_3$.

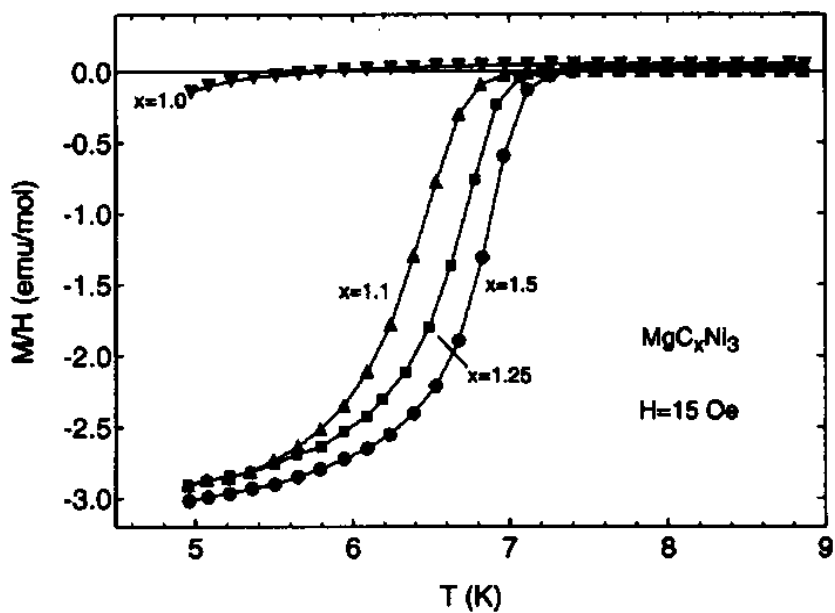

Fig. 2. Magnetic characterization of the superconducting transitions for polycrystalline powders of the intermetallic perovskite superconductor of nominal composition $MgC_xNi_3$. Applied field 15 Oe, zero field cooled data. For x=1.25 nominal composition, the perovskite composition is $MgC_{0.96}Ni_3$.

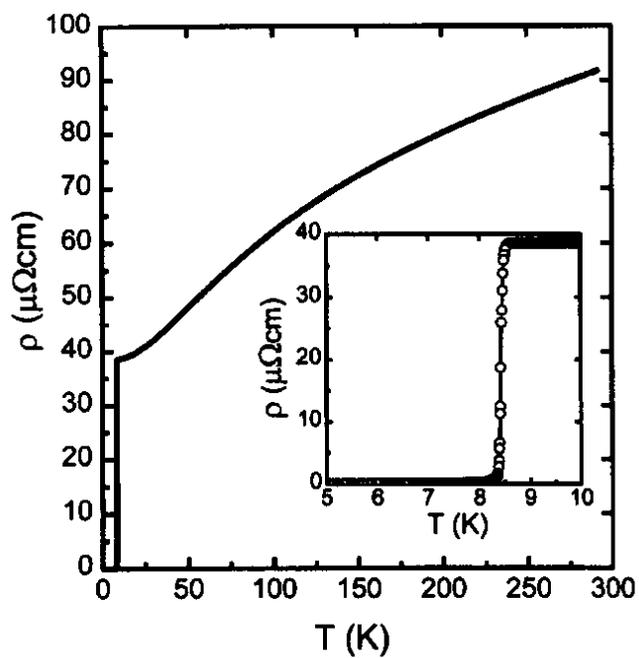

Fig. 3. The temperature dependent resistivity for polycrystalline $MgCNi_3$ between 290 and 5 K (sample of nominal composition $MgC_{1.5}Ni_3$). The normal state resistivity and a detailed view (inset) of the superconducting transition are shown. The four-probe AC method was employed, current 100mA (normal state), 10 mA (transition region), f=16.0 Hz.

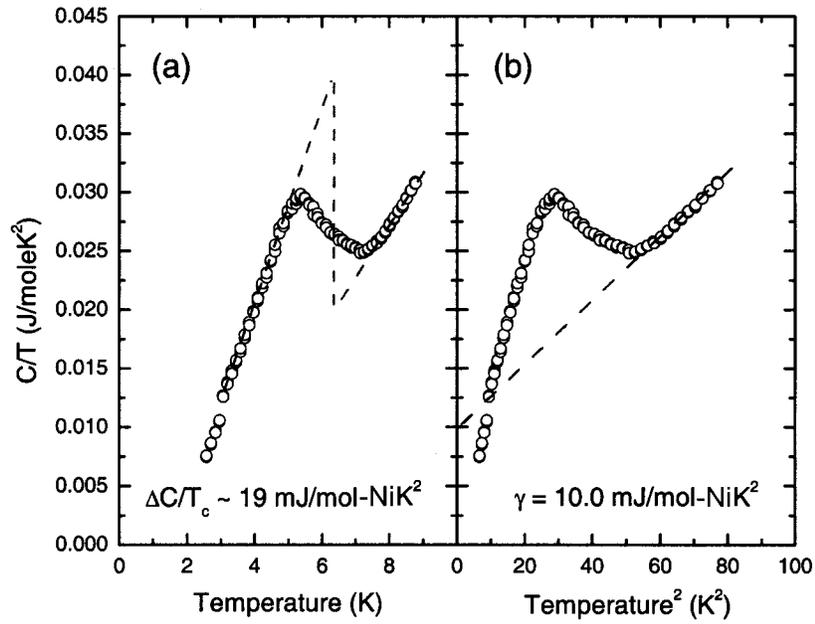

Fig. 4. The characterization of the superconducting transition in MgCNi$_3$ by specific heat. (a) C/T vs. T, allowing the determination of the change in specific heat at the superconducting transition. (b) C/T vs. T$^2$, allowing the determination of the electronic contribution to the specific heat ($\gamma$) by extrapolation to T=0.